# Tunable and Broadband Coherent Perfect Absorption by Ultrathin Black Phosphorus Metasurfaces

TIANJING GUO, CHRISTOS ARGYROPOULOS[*]

*Department of Electrical and Computer Engineering, University of Nebraska-Lincoln, Lincoln, Nebraska 68588, USA*
*[*christos.argyropoulos@unl.edu](*christos.argyropoulos@unl.edu)*

**Abstract:** Black phosphorus (BP), a relative new plasmonic two-dimensional (2D) material, offers unique photonic and electronic properties. In this work, we propose a new tunable and broadband ultrathin coherent perfect absorber (CPA) device operating in the terahertz (THz) frequency range. It is based on a bifacial metasurface made of BP patch periodic arrays separated by a thin dielectric layer. Broadband CPA bandwidth is realized due to the ultrathin thickness of the proposed device and the extraordinary properties of BP. In addition, a substantial modulation between CPA and complete transparency is achieved by adjusting the phase difference between the two counter-propagating incident waves. The CPA performance can be tuned by dynamically changing the electron doping level of BP. The CPA response under normal and oblique transverse magnetic (TM) and electric (TE) polarized incident waves is investigated. It is derived that CPA can be achieved under both incident polarizations and across a broad range of incident angles. The presented CPA device can be used in the design of tunable planar THz modulators, all-optical switches, detectors, and signal processors.

## 1. Introduction

The absorption of electromagnetic energy by a conductive material can be greatly enhanced due to interference between the incident and reflected waves [1–3]. The recently proposed concept of coherent perfect absorption (CPA) [4–7] has generalized this feature to free-standing structures based on the destructive interference of two input waves interacting inside their lossy materials. It can lead to total absorption in a flexible and controllable way by efficiently manipulating the interference effect via varying the incident waves' intensities or phases. Recently, the CPA concept has attracted increased research interest [8–14] because of its potential use in nanophotonic applications, such as optical modulators, optical switches, and sensors. Moreover, CPA devices have been realized with different versatile geometries. The use of metamaterials or metasurfaces [15–18] has opened new possibilities in the efficient CPA control through the incorporation of additional robust system parameters and artificial engineered materials. Moreover, two-dimensional (2D) materials [19], such as graphene, promise to provide further tunability in the CPA response through dynamically controlling their doping level by using appropriate gating voltage configurations [13]. This dynamic feature cannot be achieved with other systems based on conventional bulky materials.

CPA consists the time-reversed analog of laser, inevitably leading to narrow bandwidth operation [4]. However, it has been reported that the CPA bandwidth can be improved by substantially decreasing the size of the absorbing medium [20], i.e., making the CPA structure much smaller than the incident wave wavelength. This effect is impossible to be achieved by using conventional lossy materials because enormous loss coefficients will be required to attain CPA that are difficult to be realized in practice [11]. 2D materials provide a practical alternative to design ultrathin CPA devices (much smaller than the operation wavelength), because they can sustain strongly confined subwavelength plasmonic waves along their surfaces [21–26]. The resulted compactness of the CPA devices based on 2D materials is another significant advantage, besides their unique tunability feature.

Graphene is the most widely investigated 2D material due to its outstanding optical (plasmonic), mechanical, and electronic properties [27–30]. As an alternative to graphene, recently, black phosphorus (BP) demonstrated prominent potential to a variety of applications including photodetectors, phase shifters, absorbers, and field effect transistors [31–35]. It can be manufactured either by using exfoliation or other mechanical/chemical deposition techniques [36–38]. By patterning the monolayer BP to circular or rectangular patches, we can design plasmonic metasurfaces operating at THz frequencies [39–41]. In addition, unlike graphene and other 2D materials, BP exhibits a strong anisotropic plasmonic response because of the puckered honeycomb lattice structure formed by its atoms [42,43]. Another difference compared to graphene is that BP has a thickness-dependent direct bandgap, ranging from $\sim 0.3eV$ (bulk BP) to $\sim 2eV$ (monolayer BP) [44–46]. This characteristic property makes BP a prominent material to enable tunable optical response over a broad wavelength range [47,48]. Note that 2D BP monolayers have already been used in the design of conventional THz absorbers [40,49–52] under single wave illumination.

In this paper, a new tunable and broadband ultrathin THz CPA device based on two bifacial BP-based metasurfaces separated by a thin dielectric spacer layer is designed. The proposed structure is extremely thin with subwavelength thickness on the order of $\lambda_0/1500 - \lambda_0/2000$, where $\lambda_0$ is the operation wavelength. We provide the theoretical analysis of the proposed planar and compact CPA device by using the transfer matrix and equivalent circuit models. Numerical simulations are used to verify the accuracy of the theoretical models and provide further physical insights. Thanks to the properties of the BP patches, the CPA resonance can be dynamically tuned by changing the electron doping level of BP. The CPA effect can be realized for both transverse magnetic (TM) and electric (TE) polarized incident waves. Furthermore, a CPA angular tolerance as high as 60° is achieved for TM polarized illumination. Finally, we compare the CPA performance of the proposed structure with a similar graphene-based device and find that the broad bandwidth feature is a unique advantage of the presented BP-based device. The flexible tunability, wide bandwidth operation, and ultrathin thickness imply the great potential of the proposed device to be used in the design of tunable compact planar THz modulators, detectors, switches, and signal processors.

**2. Theoretical analysis and design of coherent perfect absorber**

Fig. 1(a) demonstrates the schematic of the proposed CPA device, which is composed of two BP monolayers structured in a patch array formation (bifacial metasurface) and separated by a thin dielectric spacer layer. The geometrical parameters of the unit cell are shown in Fig. 1(b). The transfer matrix $S$ formalism is used to theoretically analyze the propagation of the two incident waves and their interactions with the proposed BP-based CPA device. Thus, the output waves ($O_\pm$) can be expressed as:

$$\begin{bmatrix} O_+ \\ O_- \end{bmatrix} = S \begin{bmatrix} I_+ \\ I_- \end{bmatrix} = \begin{bmatrix} r_+ & t_- \\ t_+ & r_- \end{bmatrix} \begin{bmatrix} I_+ \\ I_- \end{bmatrix}, \quad (1)$$

where $r_+, r_-$ and $t_+, t_-$ are the reflection and transmission coefficients at the forward and backward directions, respectively. To quantitatively investigate the CPA effect, we define the output coefficient variable $\Theta$ as the ratio of the output waves intensities to that of the input waves [53,54]:

$$\Theta = \frac{|O_+|^2 + |O_-|^2}{|I_+|^2 + |I_-|^2} = \frac{|r + te^{i\Delta\varphi}|^2 + |t + re^{i\Delta\varphi}|^2}{2}, \quad (2)$$

where we utilized $r_\pm = r$ and $t_\pm = t$ due to symmetry and reciprocity. Furthermore, we assume that the amplitude of the forward and backward waves are always equal throughout this work, which means that one input source can be used that can be split in two counter-propagating waves with different phases by using a beam splitter and phase delay configuration. Hence, the relation between the two counter-propagating incident beams is: $I_+ = e^{i\Delta\varphi}I_-$, where $\Delta\varphi$ is the phase difference introduced by the different propagation distance between the incident waves or another phase delay configuration. CPA can be achieved when the output coefficient becomes $\Theta = 0$, which means that there will be no outgoing waves $(O_\pm = 0)$ and the total incident energy will be fully absorbed by the device. Quasi-CPA in the THz regime has been reported in the case of suspending BP [55] and graphene [56] monolayers, but this response was not broadband and omnidirectional. It was concluded that $|t|=|r|$ under single wave incidence is the necessary condition to achieve the CPA performance. We derive a more precise CPA condition that also takes into account the phase difference of the two incident waves. By rearranging the output coefficient formula, we obtain $\Theta = |r^2 + t^2 + 2rt\cos\Delta\varphi|$. In this case, CPA $(\Theta = 0)$ can be achieved if $r = -t$ and $\cos\Delta\varphi = 1$ $(\Delta\varphi = 2n\pi)$ or if $r = t$ and $\cos\Delta\varphi = -1$ $(\Delta\varphi = (2n+1)\pi)$, with n being an arbitrary integer number. Note that $r = -t$ implies that the amplitude of the reflection and transmission coefficients are the same and their phase difference $\Delta\theta$ (not to be confused with the phase difference $\Delta\varphi$ between the two counter-propagating waves) is $\pi$, i.e., $|r|=|t|$ and $\Delta\theta = \pi$. Similarly, $r = t$ means that $|r|=|t|$ and $\Delta\theta = 0$.

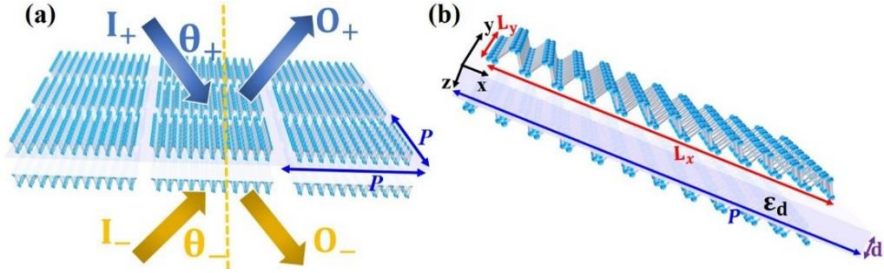

Fig.1. (a) Schematic of the BP-based CPA device consisting of two metasurfaces made of BP patches separated by a thin dielectric spacer layer. $I_\pm$ and $O_\pm$ denote the input and output waves, respectively, from each side. $\theta_+$ and $\theta_-$ are the incident angles of the forward and backward incident waves. (b) Unit cell of the proposed device. $L_x$ and $L_y$ are the lengths of each BP patch in the x- and y- direction, respectively. $P$ and $d$ are, respectively, the period of the BP patch array and the thickness of the dielectric layer. The lattice structure of monolayer BP can also be seen in this figure.

We employ the transmission-line method [57], appropriately modified to model the current BP-based design, to compute the reflection, transmission, and absorption coefficients when a single incident wave illuminates the proposed device. The complex anisotropic conductivity of a BP monolayer can be described by the semi-classical Drude model [39]: $\sigma_j = iD_j/\pi(\omega + i\eta/\hbar)$, where $D_j = \pi n_s e^2/m_j$ is the Drude weight, $j = x, y$ represents each in-plane direction, $n_s$ is the electron doping level, $e$ is the electron charge, $\omega$ is the incident radial frequency, $\hbar$ is the reduced Planck's constant, and $\eta = 10 meV$ is a typical relaxation rate value for BP [39]. The electron mass of BP along the x- (armchair) and y- (zig-zag) direction can be expressed as: $m_x = \hbar^2/(2\gamma^2/\Delta + \eta_c)$ and $m_y = \hbar^2/2\nu_c$, respectively, where $\gamma = 4a/\pi eVm$, $\Delta = 2eV$, $\eta_c = \hbar^2/(0.4m_0)$, and $\nu_c = \hbar^2/(1.4m_0)$ [42]. Here, the thickness of the BP monolayer $a$ is set equal to 0.223 nm. The surface impedance of the proposed metasurface based on an array of periodic BP patches can be expressed as [58–60]:

$$Z_{sj} = [\frac{P}{L_j \sigma_j} - i\frac{\pi}{\omega\varepsilon_0(\varepsilon_d + 1)P\ln\{\csc(\pi(P-L_j)/2P)\}}]\frac{L_j}{L_i}, \qquad (3)$$

where $j$ represents the x- (or y-) in-plane direction, and $i$ represents the y- (or x-) direction, respectively, $P$ is the period of the patch array, and $\varepsilon_d$ is the relative permittivity of the dielectric spacer layer. Eq. (3) can be simplified as:

$$Z_{sj} = \frac{(\eta/\hbar)\pi P}{L_i D_j} + i[\frac{\omega \pi P}{L_i D_j} - \frac{1}{\omega C_j}], \quad (4)$$

where $C_j$ is the effective capacitances in the $j$ direction, which is determined by the dielectric spacer layer and the geometrical parameters of the BP patch array:

$$C_j = (L_j/L_i)(1/\pi)\varepsilon_0(\varepsilon_d + 1)P\ln\{\csc(\pi(P-L_j)/2P)\}. \quad (5)$$

Eq. (4) demonstrates that the surface impedance of the BP patch array is anisotropic. In addition, it can be derived that the BP patch array can be modeled by an effective resistor (R), inductor (L), and capacitor (C) in series, as schematically shown in Fig. 2(a). The wave impedance of the dielectric spacer layer for normal incidence illumination is equal to: $Z_d = i\eta_d \tan(k_d d) = 1/(i\omega C_d)$, where $\eta_d = \sqrt{\mu_0/\varepsilon_0 \varepsilon_d}$ and $k_d = \omega\sqrt{\varepsilon_d \varepsilon_0 \mu_0}$ are the characteristic impedance and wave number, respectively, and $C_d$ is the effective capacitance that can be calculated by the $Z_d$ formula.

First, we investigate the proposed device under TM-polarized incident wave illumination. In this case, the electric field is polarized along the x-direction and the transmission line equivalent circuit model is shown in Fig. 2(a). Both the top and bottom BP patch arrays form identical R-L-C series circuits with parameters shown in Fig. 2(a). The total impedance of the device can be computed by the parallel combination of the two R-L-C series circuits and the capacitive impedance of the dielectric layer: $Z = Z_d \| 0.5Z_{sx}$, where $Z_{sx}$ is the corresponding surface impedance of the BP patch array in the x-direction due to TM-polarized illumination. Then, the input impedance of the proposed device is given by: $Z_{in} = Z_d \| 0.5Z_{sx} \| Z_0$, where $Z_0 = 120\pi \Omega$ is the characteristic impedance of the surrounding free space. As a result, the total reflection, transmission, and absorption coefficients [61–64] of the proposed CPA THz device can be computed by: $r=(Z_{in}-Z_0)/(Z_{in}+Z_0)$, $t=2Z_{in}/(Z_{in}+Z_0)$ and $A=1-|r|^2-|t|^2$, respectively. Note that a relationship exists that connects the input and surrounding free space impedance to perfectly satisfy the CPA condition: $Z_{in} = 0.5Z_0$ [65]. The proposed bifacial metasurface design can decrease the total input impedance due to mutual coupling between the BP patches and realize the aforementioned CPA impedance condition in a broad frequency range. As a result, the bifacial metasurface design is required to achieve CPA performance. By using only a single metasurface, the input impedance of the proposed device will increase $(Z_{in} > 0.5Z_0)$ and the CPA condition will not be possible to be satisfied. Hence, the capacitive coupling between the two BP-patches is the necessary condition in order to obtain efficient CPA [66,67].

The parameters used for the thin dielectric spacer layer are $\varepsilon_d = 2.92$ and $d=20nm$. The periodicity of the device is $P=500nm$. The electron doping level of the BP monolayer is chosen to have a moderate value of $n_s=4\times10^{13}cm^{-2}$. The side lengths of each BP patch in the x- and y-directions are equal: $L_x= L_y=480nm$. The spectra of the reflection, transmission, and absorption coefficients computed by using the equivalent circuit model are demonstrated in Fig. 2(b). It can be seen that $|r|=|t|$ at the resonant frequency (8 THz). In addition, the absorption reaches to 50% at this resonance point, which satisfies the necessary condition to achieve CPA. In order to quantitatively demonstrate the CPA effect, we analytically calculate the output coefficient $\Theta$ by substituting the computed reflection and transmission coefficients into Eq. (2). Indeed, CPA can be achieved at the resonant frequency 8 THz when the phase difference between the two counter-propagating incident waves is $\Delta\varphi = 2n\pi$, as it is shown in Fig. 2(c). The absorption can reach to 90% values at the off-resonant frequencies of 7.2 THz and 8.8 THz, which implies broadband quasi-CPA performance. This is typical bandwidth convention used to characterize the frequency response of numerous CPA systems [68]. While complete absorption cannot be achieved everywhere in this frequency range, even a 90% absorption contrast can be very useful for several interesting applications, such as all-optical switches, sensors, absorbers [68]. The proposed structure achieves an ultrabroadband bandwidth performance of 1.6 THz where 90% absorption is achieved. It is worth noting that 99% absorption can also be realized within a broad bandwidth of 0.4 THz. The pink dotted line in Fig. 2(c) demonstrates the performance of the proposed CPA device for higher frequencies (10 THz), where the CPA effect ceases to exist. Note that the CPA can be modulated from 0.01 % to 99.98 % at the resonant frequency (8 THz) just by varying the phase difference $\Delta\varphi$ between the counter-propagating incident waves. This means that the proposed planar compact device can be switched from perfect absorption to complete transparency. In practical applications, a set of different designs of lossless Mach-Zehnder (MZ) interferometers [7,20] can be used to realize the specific phase modulation value for each narrow frequency range of the total broadband operation. However, this experimental set-up will be very complex and will require a plethora of additional input optical components to be implemented. A simpler alternative can be a lossy broadband beam splitter, as it was demonstrated in [20]. The lossy broadband beam splitter component can be a thin metallic film at Woltersdorff thickness [69]. There are also several other approaches to realize broadband beam splitters, such as by utilizing matched dispersion in transmission and reflection modes [70], evanescent coupling based on an asymmetric directional coupler [71], and by using the inherent anisotropy and dispersion of sub-wavelength metamaterials [72].

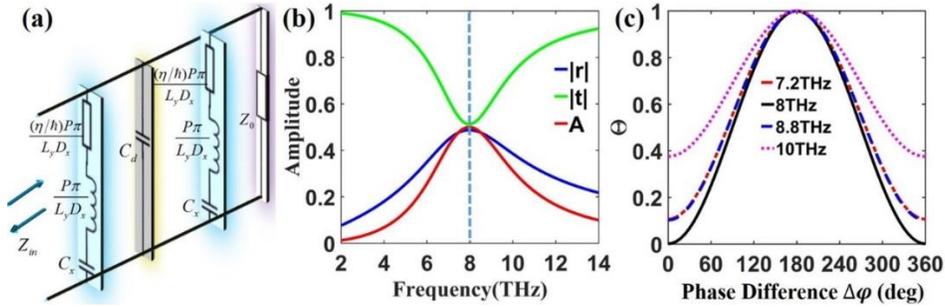

Fig. 2. (a) Equivalent circuit model of the proposed BP-based CPA device under a TM-polarized incident excitation. (b) The reflection, transmission, and absorption coefficients spectra under a single TM-polarized wave illumination. (c) Output coefficient Θ as a function of the phase difference between the two incident counter-propagating beams at four different frequencies.

## 3. Results and discussion

The proposed CPA device is further investigated and optimized by performing simulations with COMSOL Multiphysics [73], a commercial electromagnetic solver based on the finite element method. The use of a 3D simulation domain is necessary because the BP monolayer is modeled as a boundary layer with an anisotropic surface current distribution. The BP optical conductivity was given in Section 2 and this formula is used in the simulations. Periodic boundary conditions (PBC) are set in the x- and y-directions to model the periodic BP patches and port boundaries are placed in the z-direction to create the incident plane waves. Considering that BP is highly reactive to the oxygen of the surrounding free space, in a practical scenario, it is essential to encapsulate the entire device inside an ultrathin dielectric sheet to protect the BP monolayers from degradation [74]. Thus, a 10 nm thick $Al_2O_3$ thin film is used on both sides of the device that now can be exposed to the surrounding air. In a potential experimental implementation of the proposed CPA device, first, bulk BP can be purchased from commercial vendors. Then, the fabrication can start with the micromechanical cleavage of bulk BP crystals directly onto the $Al_2O_3$ substrate from both sides, which will form the dielectric spacer layer between the metasurfaces. The 10 nm thick $Al_2O_3$ protective thin film will be grown on both metasurface sides by atomic layer deposition [52,75]. These thin dielectric films will be used as encapsulating layers to protect the BP monolayer from degradation. The structural characterization of BP can be performed with the help of atomic force microscopy (AFM) and optical microscopy to make BP resonate at the certain frequency of choice [75].

First, we consider the case of a single normal incident TM-polarized THz wave impinging on the device. The simulated spectra of transmission, reflection, and absorption coefficients, and the phase difference $\Delta\theta$ between the transmission and reflection coefficients are shown in Fig. 3(a). In these results, the same parameters were chosen for the system with those used in the equivalent circuit model presented in Section 2. We will keep using these parameters during this work unless otherwise specified. It can be clearly seen in Fig. 3(a) that the phase difference between the transmission and reflection coefficients is $\Delta\theta = \pi$ and the transmission $|t|$ and reflection $|r|$ amplitudes are very close to each other at the peak absorption resonance frequency (8.4 THz). According to the theoretical analysis, the CPA condition $r \approx -t \approx 0.5$ is perfectly satisfied at this point. Note that the simulation results shown in Fig. 3(a) agree very well with the theoretical results presented in Fig. 2(b), despite that in the simulations the encapsulation of BP with a thin dielectric layer was also included. This dielectric layer is extremely thin compared to the wavelength of the operating THz frequency. The small frequency shift between the theoretical and simulation results can be attributed to both the finite mesh size used during the simulations and the thin dielectric layer. The broadband absorption response $(A = 0.5)$ under a single beam illumination shown in Fig. 3(a) implies the potential to accomplish broadband CPA performance, since the CPA condition is satisfied for a relative broad frequency range. The proposed device will function as a deep-subwavelength CPA ultrathin film. In order to further verify the CPA formation condition, we calculate the effective surface conductivity $\sigma_\parallel^e$ by using a retrieval method adapted to thin films [76]. Fig. 3(b) presents the real, imaginary, and absolute values of the effective surface conductivity of the proposed structure normalized to the free space admittance $Y_0 = 1/Z_0$. It is verified that the CPA condition $|\sigma_\parallel^e|Z_0 = 2$ [65,68,77] is fulfilled around the resonant frequency, which coincides with the 50% absorption maximum [62,78].

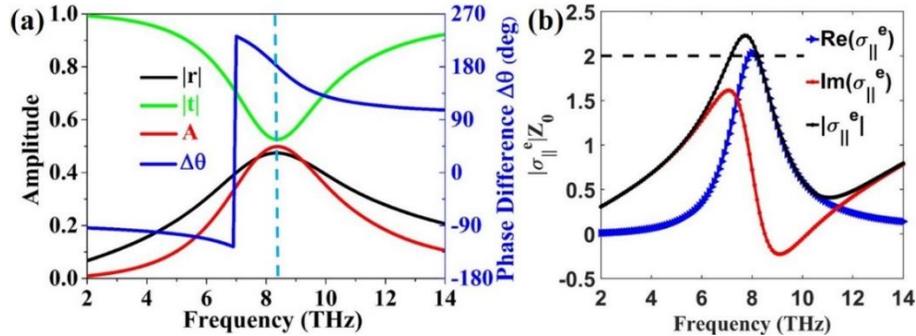

Fig. 3. (a) Transmission, reflection, and absorption coefficients, and the phase difference between the transmission and reflection coefficients, as a function of frequency for a single TM-polarized wave normally incident upon the CPA device with electron doping level of $4\times10^{13}cm^{-2}$. (b) Real, imaginary, and absolute values of effective surface conductivity $\sigma_\parallel^e$ of the proposed BP-based CPA device normalized to the free space admittance.

Similar to the theoretical analysis presented in Section 2, we launch a second incident wave to the device from the opposite side that has the same intensity and polarization with the first one. Then, the two counter-propagating waves destructively interfere at the nanoscale thickness of the proposed metasurface leading to CPA formation. To further investigate the CPA process and confirm its dependence on the phase difference $\Delta\varphi$ between the incident waves, we create a contour plot of the computed output coefficient $\Theta$ as a function of the incident frequency $f$ and phase difference $\Delta\varphi$ that is shown in Fig. 4(a). Interestingly, the total absorption can reach to 100 % (CPA, $\Theta=0$) with a proper phase difference of $\Delta\varphi=2n\pi$, while it is almost 0 % (complete transparency, $\Theta=1$) with $\Delta\varphi=(2n+1)\pi$.

The output coefficient $\Theta$ as a function of the phase difference at four different frequencies is shown in Fig. 4(b). It is important to notice that the simulation results agree well with our analytical calculations in Fig. 2(c). In order to quantitatively measure the tunable performance of the CPA response, we define the CPA modulation contrast as the ratio of the maximum to the minimum output coefficient $\Theta$ value. The CPA modulation contrast can reach very high values of approximately 54 dB at 8.4 THz, outperforming the modulation contrast predicted in [79] for a different graphene-based CPA device. Furthermore, we calculate the CPA modulation contrast at 7.7 THz and 9.2 THz, which is 21.9 dB and 20 dB, respectively. For comparison, the pink dotted line in Fig. 4(b) is the output coefficient versus the phase difference at a non-resonant frequency (10 THz), where no CPA is achieved. Clearly, Figs. 4(a) and 4(b) demonstrate that almost perfect absorption can be obtained within a wide frequency range when the phase difference is fixed to $\Delta\varphi=2n\pi$.

The broad bandwidth feature is a major advantage of the proposed CPA device. It can be attributed to its extremely thin planar geometry, since it is known that the CPA bandwidth is inversely proportional to the CPA film thickness [65], combined with the unique dispersive properties of BP (see Fig. 8 later in the manuscript). The proposed device acts as a deeply subwavelength CPA ultrathin film with a total thickness much smaller than the operation wavelength $\lambda_0$, on the order of $\lambda_0/1500 - \lambda_0/2000$. The physical mechanism behind the broadband CPA phenomenon is that the tangential electric field component is almost constant across the presented CPA device due to the structure's ultrathin thickness. This leads to a nearly unity normalized electric field distribution inside the ultrathin structure, combined with a magnetic field distribution exhibiting a dip at its center (in the middle of the dielectric layer), causing perfect CPA in a broad frequency range. In the current case, this effect is possible without resorting to the use of narrowband reflectors terminating the structure that limit the absorption bandwidth [80]. Note that extremely thin and non-resonating conventional materials, different from the currently used BP, will require enormous and non-practical loss coefficients in order to achieve CPA with such a nanoscale geometry [11].

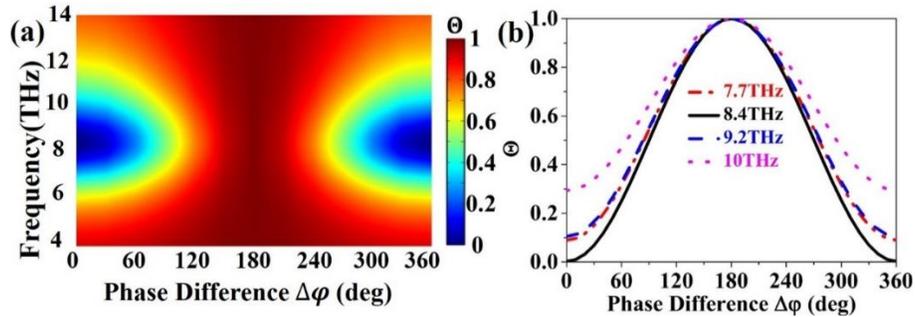

Fig. 4. (a) Contour plot of the output coefficient Θ as a function of frequency *f* and phase difference Δφ between the two counter-propagating incident waves. (b) Output coefficient Θ as a function of the phase difference at four different frequencies. Perfect CPA is obtained at 8.4 THz for Δφ = 2*n*π.

Next, we consider how the CPA response of the proposed BP-based device varies with its geometrical features. The dependence of the output coefficient on the thickness of the dielectric spacer layer under TM-polarized normal incident counter propagating waves is shown in Fig.

5(a). It is worth noting that broadband CPA can be achieved under a wide range of thickness values. This is really crucial to the practical implementation of the proposed device, since the thickness of the ultrathin dielectric spacer layer might be slightly changed during the fabrication process. The CPA frequency redshifts as we gradually increase the thickness of the dielectric layer from 1nm to 20 nm and the result can be seen in Fig. 5(a). The CPA frequency redshift becomes minor for thicknesses larger than 20 nm. For practical reasons, we assume the dielectric layer thickness to be 20 nm throughout this manuscript. The absorption resonance can be more strongly affected by the size of the BP patches [39,51]. Note that the thickness of the proposed structure can in principle be as low as 1 nm, which is a much lower value compared to previously proposed bulky structures investigated to achieve broadband CPA with thicknesses comparable or even larger to the wavelength of the incident radiation [20,81–83]. Hence, the currently presented structure uniquely achieves a broadband CPA response with a much more compact and subwavelength thin configuration. The only work that can compare to the results presented in the current paper is [19], where the authors have demonstrated an ultrathin broadband nearly perfect THz CPA device by heavily doping a silicon film. However, the lowest thickness they reached with their design is 150 nm by using a very high and not practical doping level value, which is still much thicker compared to the thickness of our proposed BP metasurface that is approximately equal to 20 nm and can become even thinner, as it is shown in Fig. 5(a). Furthermore, the currently presented CPA design has several additional advantages compared to [16] because it can operate under different polarized incident waves and for a broad range of incident wave angles, as it will be shown later in the paper.

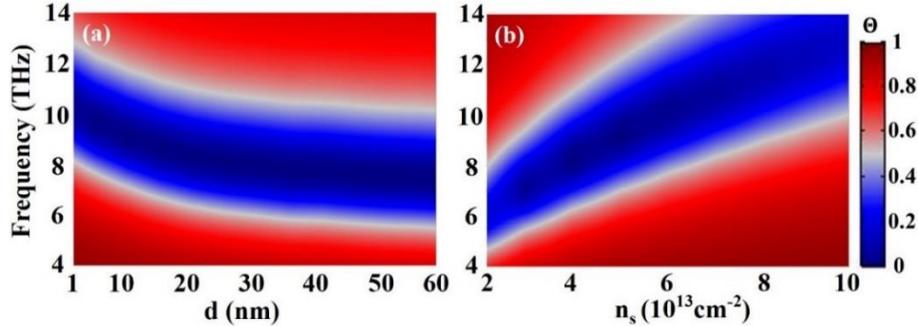

Fig. 5. Contour plot of the computed output coefficient $\Theta$ as a function of the frequency $f$ of the TM-polarized incident waves and (a) the thickness of the dielectric spacer layer or (b) the BP electron doping level $n_s$. The two counter-propagating incident beams have a fixed phase difference equal to $\Delta\varphi = 2n\pi$.

It is interesting to investigate the effect of the electron doping level on the CPA performance, since it can alter the surface conductivity of the BP monolayer. This interesting tunable characteristic has been employed for optical modulator [84] and photodetector [85] applications. The change in the CPA performance by tuning the electron doping level of the BP monolayer between moderate practical values from $2\times10^{13}\ cm^{-2}$ to $1\times10^{14}\ cm^{-2}$ [86] is demonstrated in Fig. 5(b). The coherent absorption is kept nearly perfect, as long as the electron doping level $n_s$ is higher than $3\times10^{13}\ cm^{-2}$. In addition, the CPA resonance frequency blue shifts as the electron doping level of the BP monolayers increases. This is because the electron doping values can significantly alter the BP properties. An approximate relationship between the resonance frequency and the electron doping level is $f \propto \sqrt{n/L}$ [42,51], where $n$ is the electron doping level and L is the length of the BP patch. Clearly, this relationship is consistent with the simulation results shown in Fig. 5(b). The electron doping level also has a substantial influence on the bandwidth of the CPA effect, as shown in Fig. 5(b), which is attributed to the slower variation of the BP patches' surface impedances at higher frequencies. Thus, the CPA effect can be flexibly tuned by dynamically changing the electron doping level values of each BP patch.

Up to now we have considered only the CPA performance with TM-polarized incident waves, similar to several previous works relevant to plasmonic CPA devices [68,87,88]. Next, we investigate the CPA performance of the proposed device under TE-polarized incident beams. In this case, the electric field of the incident plane waves is polarized along the zigzag (y) direction of the BP patches. The same CPA structure with identical geometry to the previously studied TM-polarized illumination case is investigated. First, a single incident wave is launched. The computed reflection and transmission coefficients are not equal but very close to each other at the resonance frequency of 8 THz, as it is shown in Fig. 6(a). However, the computed absorption can reach to 50%, which is the most vital and prevalent result in order to achieve CPA. In addition, the phase difference between reflection and transmission coefficients is exactly 180° at this resonant frequency, indicating that the CPA response can also be achieved for TE-polarized incident waves.

Subsequently, we launch a second beam with the same TE polarization but from the opposite side. The tunability of the proposed CPA device under TE-polarized excitation is investigated in Fig. 6(b). CPA $(\Theta = 0)$ can be achieved when the electron doping level is larger than $7 \times 10^{13} cm^{-2}$, which is slightly increased values compared to the TM-polarization case, because BP has increased losses along its zigzag direction [39]. Based on this property, it has been shown recently that BP slabs can be used to manipulate the polarization of the incident light [89]. Note that the increase of the electron doping level in the zigzag direction under the TE-polarized excitation has the same effect with the TM-polarized case, leading to blue shift in resonance frequency and broader CPA bandwidth. We emphasize that CPA can be achieved under both of TM- and TE-polarized illuminations with the currently proposed configuration, but under different doping levels. In particular, the CPA response can be achieved for lower doping values under TM-polarized radiation and for higher doping level with TE-polarized excitation, leading to a tunable CPA response based on the incident polarized radiation and BP doping level. The CPA bandwidth is broader for TM polarization [Fig. 5(b)], compared to TE [Fig. 6(b)], in the case of low doping values. For higher doping levels, the CPA is broadband for both polarizations but its resonant frequency is stronger blueshifted only for the TM-polarized illumination case. Hence, BP shows strong anisotropic characteristics in armchair (x) and zigzag (y) directions and its electron doping level can be tuned, making it possible to attain CPA for both polarizations and different doping values.

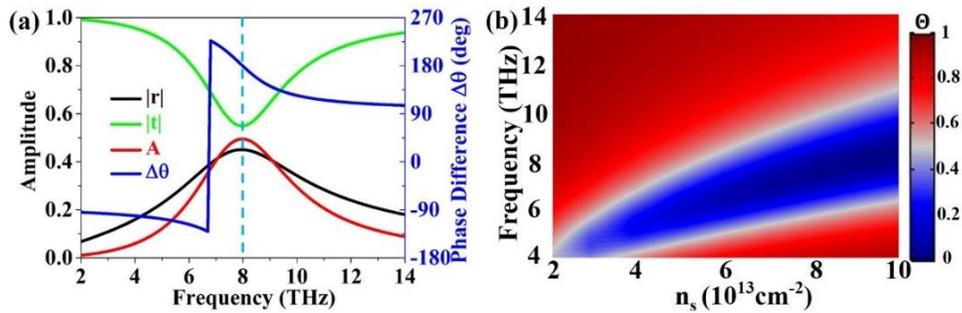

Fig. 6. (a) Transmission, reflection, and absorption coefficients, and the phase difference between the transmission and reflection coefficients, as a function of frequency for a single TE-polarized wave normally incident upon the CPA device with electron doping level of $9 \times 10^{13} cm^{-2}$. (b) Contour plot of the computed output coefficient $\Theta$ as a function of the frequency $f$ and the electron doping level $n_s$. The two counter-propagating TE-polarized incident waves have a fixed phase difference equal to $\Delta\varphi = 2n\pi$.

We further study the CPA performance of the proposed device, now under oblique TM-polarized incident waves. The simulated spectra of the transmission, reflection, and absorption coefficients, and the phase difference between the transmission and reflection coefficients, under a single incident wave with an angle of 60° are shown in Fig. 7(a). The reflection and transmission coefficients are equal to each other at the resonance frequency of 9.4 THz, which is slightly shifted due to the oblique incident illumination. Meanwhile, their phase difference is exactly 180° at this point. Clearly, Fig. 7(a) demonstrates that the requirement to achieve CPA can be fulfilled with the proposed device even under 60° oblique incidence. In order to fully investigate the range of incident angles over which CPA can be attained, we compute the output coefficient as a function of the incident angle ranging from 0° to 89° and the frequency varying from 4 to 14 THz. The computed contour plot is presented in Fig. 7(b), where the omnidirectional CPA response of the proposed device is demonstrated. CPA can be achieved under incident angles as high as 60°. Omnidirectional CPA operation is another unique and very important feature of the proposed planar CPA device.

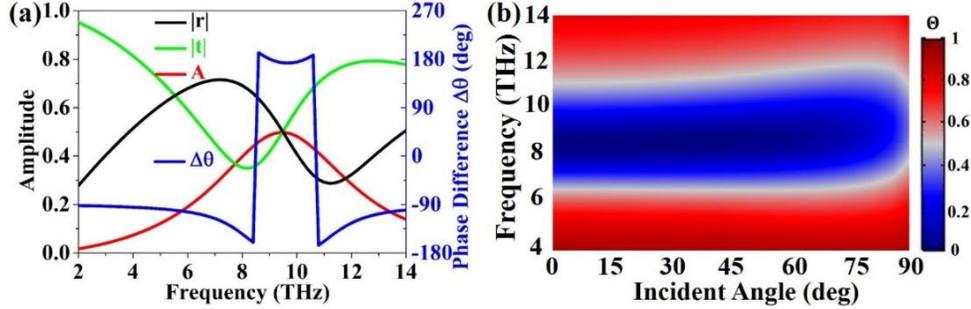

Fig. 7. (a) Transmission, reflection, and absorption coefficients, and the phase difference between the transmission and reflection coefficients, as a function of frequency $f$ for a single TM-polarized incident wave upon the CPA device with an angle of 60°. (b) Contour plot of the computed output coefficient $\Theta$ as a function of frequency $f$ and incident angle in the case of two counter propagating incident waves with a fixed phase difference $\Delta\varphi = 2n\pi$.

Finally, it worth mentioning that a similar bifacial metasurface design made of graphene is also expected to exhibit CPA performance. Graphene is the most widely-used 2D plasmonic material and has been studied in a variety of THz applications [90,91]. The envisioned graphene-based CPA device is designed to resonate around 8 THz under TM-polarized incident waves just by slightly changing the dimensions of the presented structure. The lengths of each graphene patch and the period of the nanopatch array are changed to be $L_x=L_y=360nm$, and $P=400nm$, respectively. In addition, the conductivity of graphene is described by the Drude model [92] and its Fermi energy is set to be equal to 0.2 eV. We compute the output coefficient of the graphene metasurface and compare its CPA performance with the BP-based CPA device in Fig. 8. More than 90% coherent absorption can be obtained over a broad bandwidth (1.5 THz) only in the case of the BP-based CPA device, while a narrower bandwidth (0.5 THz) is achieved with the graphene-based CPA device. The bandwidth of the proposed BP-based device is three times broader compared to the graphene-based structure. Note that we have also investigated the CPA response of the graphene-based device under different Fermi levels (not shown here) and found that the Fermi level does not alter its CPA bandwidth, as it was also derived in [39]. The broad bandwidth CPA response of the proposed BP-based device results from the unique plasmonic properties of BP.

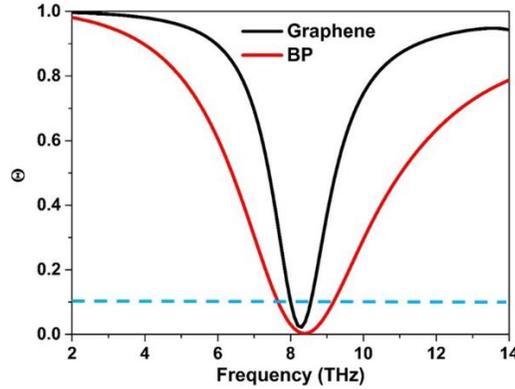

Fig. 8. Output coefficient Θ as a function of the frequency *f* under two incident counter-propagating TM-polarized waves with the same intensity and phase for the BP- (red line) and graphene- (black line) based CPA devices.

## 4. Conclusions

To conclude, we designed a new tunable and broadband THz CPA bifacial metasurface based on two metasurfaces composed of periodically arranged identical BP patches separated by a thin dielectric layer. We theoretically analyzed the proposed ultrathin CPA device based on an equivalent circuit model and then verified its performance by full-wave simulations. By calculating the output coefficient of the structure under two counter-propagating waves, we proved that broadband CPA can be achieved by using this configuration. The CPA condition can be controlled by varying the phase difference between the two incident waves. In addition, the CPA frequency point can be tuned by dynamically changing the doping level of BP. Interestingly, the tunable CPA effect can be realized for both TE- and TM-polarized illumination and achieved over a wide range of incident angles, leading to a unique omnidirectional CPA performance. The CPA bandwidth of the proposed device is much broader compared to a similar graphene-based architecture structure.

The proposed CPA structure has several unique features that cannot be found in other CPA devices proposed in the literature: 1) It is extremely thin with subwavelength thickness on the order of $\lambda_0/1500 - \lambda_0/2000$, where $\lambda_0$ is the operation wavelength. Moreover, the proposed device can work as CPA even when the thickness of the dielectric layer approaches extremely small values (1 nm), as it is proven in Fig. 5(a). The ultrathin dielectric layer, combined with the two black phosphorus one-atom-thick patch arrays, lead to an extremely thin CPA design. 2) In addition, it can have broadband operation, much broader compared to a similar structure based on graphene, as it can be seen in Fig. 8. This response is due to the black phosphorus plasmonic properties that are different compared to the graphene's dispersive properties. 3) Moreover, the presented CPA response is tunable and can be achieved by illuminating the structure under different polarized incident waves or for different doping levels. 4) Finally, the proposed CPA device can operate efficiently for a broad range of incident wave angles, as it can be seen in Fig. 7(b). All these unique advantages, combined in a single ultrathin bifacial metasurface design, make the proposed structure an ideal candidate to be used in the design of planar THz modulators, switches, detectors, and signal processors. We believe that the presented theoretical results will stimulate the interest of several experimental groups that will fabricate the proposed CPA structure and measure its unique THz CPA response.

**Funding.** National Science Foundation (Grant No. DMR-1709612); National Science Foundation Nebraska Materials Research Science and Engineering Center (Grant No. DMR-1420645); Office of Naval Research Young Investigator Program (ONR-YIP) Award (Grant No. N00014-19-1- 2384); National Science Foundation Nebraska-EPSCoR (Grant No. OIA-1557417).